# Numerical Reconstruction of the Linac Beam De-bunching in the DC-operated Booster

Xi Yang

*Fermi National Accelerator Laboratory*

Box 500, Batavia IL 60510

**Abstract**

It is difficult for us to measure the Booster ring impedance up to the GHz range due to the instrumentation limit.  Since the de-bunching process in the Linac to Booster transfer is determined by the complex impedance of the Booster ring, one can obtain an impedance model, which has the best match to the machine impedance, just by searching the optimal fit between the simulated and measured turn-by-turn resistive wall signal (RWS).

## Introduction

The complex ring impedance of Booster is the major factor, which determines the de-bunching process of the 200-MHz micro-bunch (MB) in the Linac to Booster transfer.  It is important for us to develop a numerical tool, which can be used to simulate the turn-by-turn RWS when the MB is de-bunching in the DC-operated Booster.  Here, we assume that the time structure of the MB and the ring impedance are known.  Based upon these assumptions, the de-bunching process can be simulated and expressed via the turn-by-turn RWS.  One can find the most suitable impedance model for Booster by searching the optimal fit between the simulated and measured turn-by-turn RWS.

## Method

There are two major contributions from the ring impedance to the de-bunching process. First, the real part of the ring impedance makes a charged particle lose energy according to Ohm's law.  Secondly, the complex angle of the impedance, which includes the space charge impedance, causes a phase shift to its corresponding current component.

It is unavoidable for us to describe a charged particle in the frequency domain instead of the time domain, since the impedance is expressed as a function of frequency.[1] A charged particle in the 200-MHz MB with a gaussian distribution can be treated as a *probability wave*, and this probability wave has exactly the same time structure with the MB, except *the integration of the one-particle probability wave is set to one* due to the consideration of normalization. The one-particle probability wave in the time domain is shown by eq.1.

$$i(t) = \frac{\exp\left(-\frac{t^2}{2\cdot\sigma^2}\right)}{\int_{-\tau/2}^{\tau/2}\left(\exp\left(-\frac{t^2}{2\cdot\sigma^2}\right)\right)dt}. \qquad 1$$

Here, $\tau=5$ ns is the period of the 200-MHz MB. A charged particle can be expressed in the frequency domain by the fast Fourier transform (FFT) of eq.1.

The transmission line model is used for the calculation of the ring impedance from 96 magnets in Booster.[2] The impedance R / L circuit with 37 kΩ / 40 μH matches the imaginary impedance, as shown by eq.2(a).

$$Z_{2a}(f) = \frac{R\cdot\left(\frac{2\pi f}{(R/L_2)}\right)}{1+\left(\frac{2\pi f}{(R/L_2)}\right)^2}. \qquad 2(a)$$

Here $R = 37000\Omega, L_2 = 40\mu H$.

The impedance with 37 kΩ / 100 μH matches the real impedance, as shown by eq.2(b).

$$Z_1(f) = \text{Re}(Z(f)) = \frac{R\cdot\left(\frac{2\pi f}{(R/L_1)}\right)^2}{1+\left(\frac{2\pi f}{(R/L_1)}\right)^2}. \qquad 2(b)$$

Here $R = 37000\Omega, L_1 = 100\mu H$.

The space charge impedance is calculated using eq.2(c).[3]

$$Z_{sp}(f) = -\frac{(377\cdot g\cdot f)}{(2\cdot\beta\cdot\gamma^2)}. \qquad 2(c)$$

Here, $g=3$ is the geometric factor.[3] $\beta$ and $\gamma$ are Lorentz's relativistic factors, and they are 0.713 and 1.43 at the Booster injection. Finally, the imaginary impedance is the summation of eq.2(a) and eq.2(c), as shown in eq.2(d).

$$\text{Im}(Z(f)) = Z_{2a}(f) + Z_{sp}(f). \qquad 2(d)$$

The energy loss per Booster turn ($\Delta E(f)$) from a charged particle at the current component of $f$ is calculated using eq.3(a).

$$\Delta E(f) = I(f) \cdot Z_1(f).  \qquad \text{3(a)}$$

Since $I(f)$ is the FFT of the probability wave $i(t)$ of a charged particle, the unit of $\Delta E(f)$ is eV. The phase shift per Booster turn ($\Delta\phi_e(f)$), which is caused by the revolution-period change because of $\Delta E(f)$, is calculated using eq.3(b).[4]

$$\Delta\phi_e(f) = 2\pi \cdot \frac{\Delta T(f)}{T(f)}$$
$$= 2\pi \cdot \left\{ \frac{\left[\eta\left(\frac{\Delta p(f)}{p_0}\right) \cdot T_0\right]}{(1/f)} \right\} \qquad \text{3(b)}$$
$$= 2\pi \cdot \left\{ \frac{\left[\eta\left(\frac{\Delta E(f)}{E_0}\left(\frac{1}{1+\gamma^{-1}}\right)\right) \cdot T_0\right]}{(1/f)} \right\}.$$

Here, the phase slip factor $\eta$ at the injection is 0.458, the kinetic energy $E_0$ is $4\times10^8$ eV, the revolution period $T_0$ is $2.22\times10^{-6}$ s, $T(f)=1/f$ is the period of the angular frequency $f$. The phase shift per Booster turn ($\Delta\phi_c(f)$), which is caused by the complex angle of the impedance, is calculated using eq.3(c).

$$\Delta\phi_c(f) = \tan^{-1}\left(\frac{\operatorname{Im}Z(f)}{\operatorname{Re}Z(f)}\right). \qquad \text{3(c)}$$

The total phase change per Booster turn ($\Delta\phi(f)$) by a charged particle at the current component $f$ can be calculated using eq.3(d).

$$\Delta\phi(f) = \Delta\phi_e(f) + \Delta\phi_c(f). \qquad \text{3(d)}$$

Since the 200-MHz MB de-bunches within tens of microseconds, the amplitude for each frequency component of the charged-particle probability wave can be treated as a constant. Till now, we can calculate the turn-by-turn data for all the frequency components of a charged particle probability wave, and their inverse FFT gives the turn-by-turn RWS.

## Simulation and Results

The cutoff frequency ($f_c$) was chosen to be 3.2 GHz.[5] A simple impedance model,[5] as shown in eq.4, is used in our calculation.

$$Z(f) = \text{Re}(Z(f)) + \text{Im}(Z(f))$$
$$= Z(f) \quad \text{when } f < f_c,$$
$$= Z(f) \cdot \left(\frac{f_c}{f}\right)^{3/2} \quad \text{when } f > f_c.$$

The first simulation was done at the situation of $\sigma$=0.02 ns in eq.1.[6]  The MB is injected at turn #0, as shown in Fig. 1(a).  The energy loss per Booster turn due to the real impedance is shown in Fig. 1(b).  In Fig. 1(c), phase shifts due to the real impedance, the complex angle, and the total phase shift per Booster turn are shown as the red, black, and blue curves separately.  The RWS of the bunch center is shown at turns #1, #2, #3, #5, and #10 by Figs. 1(d)-(h) separately.

Another simulation was done at $\sigma$=0.2 ns.  The MB is injected at turn #0, as shown in Fig. 2(a).  The energy loss per Booster turn due to the real impedance is shown in Fig. 2(b).  In Fig. 2(c), phase shifts due to the real impedance, the complex angle, and the total phase shift per Booster turn are shown as the red, black, and blue curves separately.  The RWS of the bunch center is shown at turns #1, #2, #3, #5, and #10 by Figs. 2(d)-(h) separately.

**Comment**

In the above calculation, the momentum of the MB in Booster is treated as a constant.  The contribution to the de-bunching of the MB from its momentum spread should be included in the future model.  Also, the cutoff frequency and the impedance model used in the calculation are borrowed from the recycler in Fermilab.  Booster might have a quite different situation.  In the future, the attempt of matching the simulation with the measurement could be an important method for making correct choices of the cutoff frequency and impedance model for Booster, etc.  Furthermore, the impedance from Booster RF cavities should be considered in the future calculation for the purpose of making the model more close to the machine condition.  Finally, the program was made using MATLAB with its built-in functions, such as FFT and IFFT, and more efforts should be investigated in the future program for a better accuracy.

## Acknowledgement

The author is especially grateful for Chuck Ankenbrandt's suggestion about how to improve the model in the future. Also thanks Dr. K. Y. Ng for his useful discussion.

**References:**

[1] K. Y. Ng, "Coupling Impedances of Laminated Magnets", FERMILAB-FN-0744.

[2] J. L. Crisp and B. J. Fellenz, "Measured Longitudinal Beam Impedance of Booster Gradient Magnets", FERMILAB-TM-2145.

[3] J. A. MacLachlan, "Wakefields and Space Charge in ESME", http://www-bd.fnal.gov/pdriver/booster/meetings.html. (2001)

[4] S. Y. Lee, Accelerator Physics. (1999)

[5] K. Y. Ng and J. Marriner, "Energy Loss of a Coasting Beam inside the Recycler Ring", FERMILAB-FN-0740.

[6] E. McCrory, "Linac Beam Bunch Length Measurements", http://www-bd.fnal.gov/pdriver/booster/meetings.html. (2003)

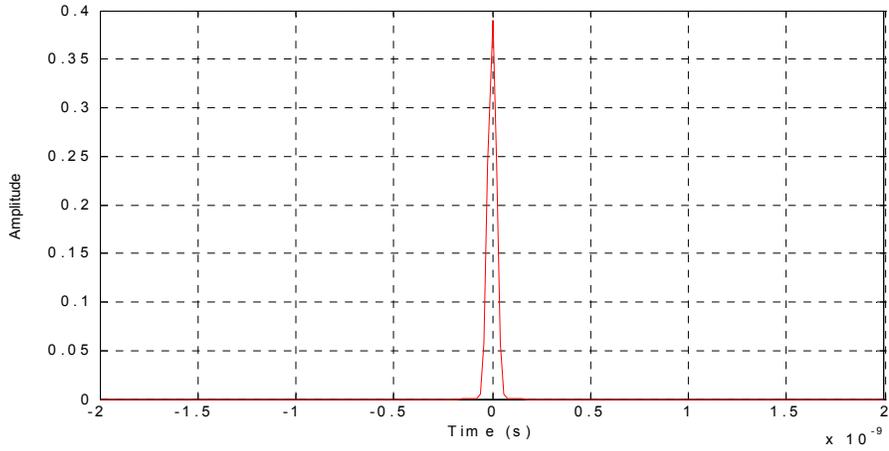

Fig. 1(a)

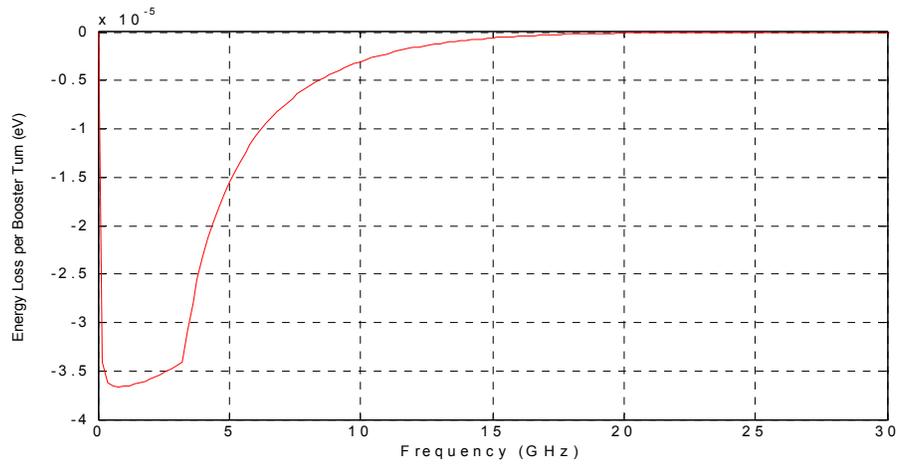

Fig. 1(b)

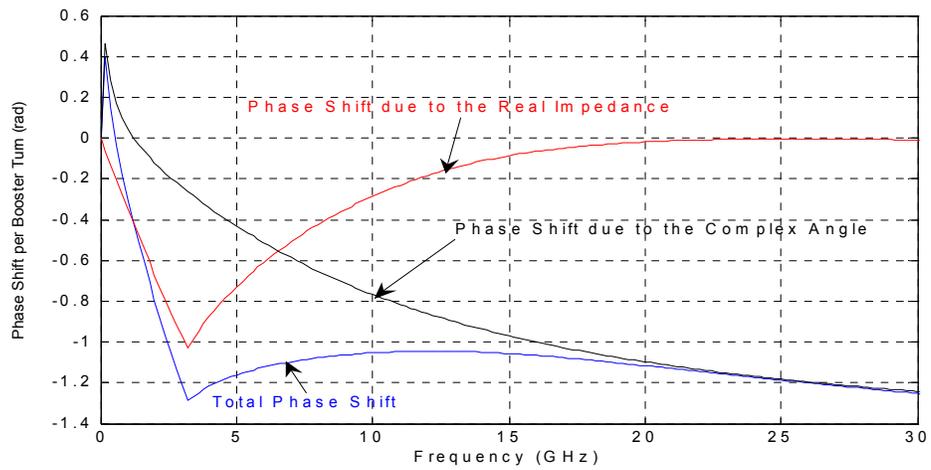

Fig. 1(c)

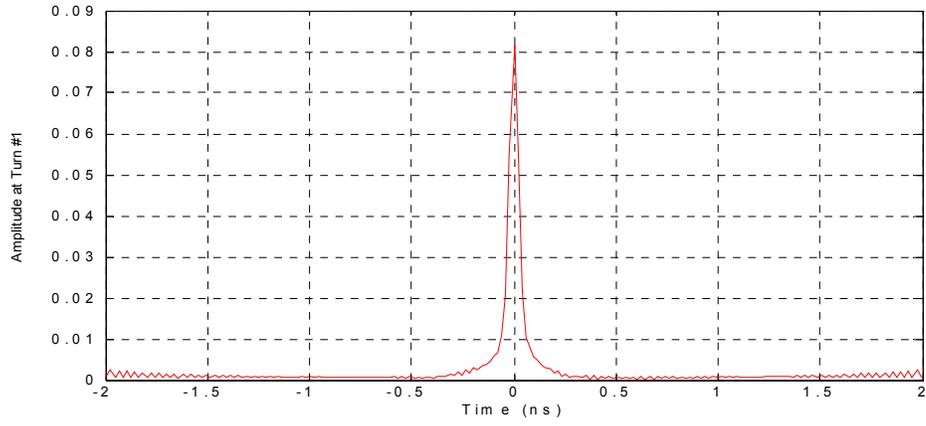

Fig. 1(d)

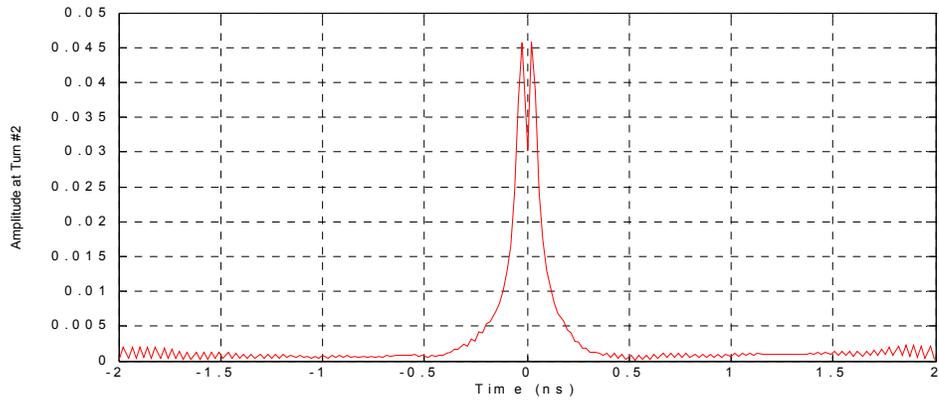

Fig. 1(e)

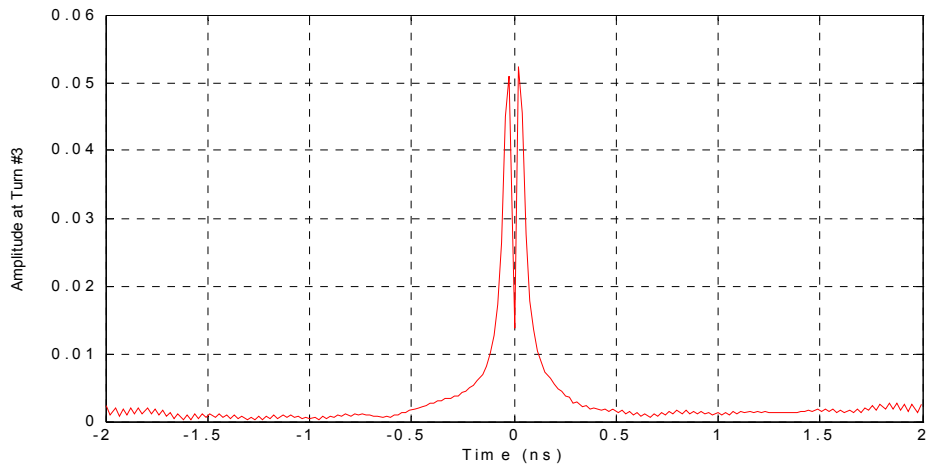

Fig. 1(f)

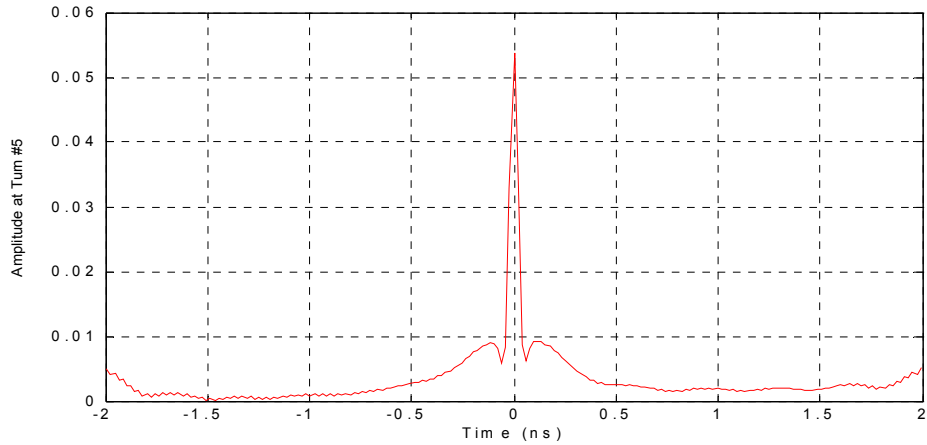

Fig. 1(g)

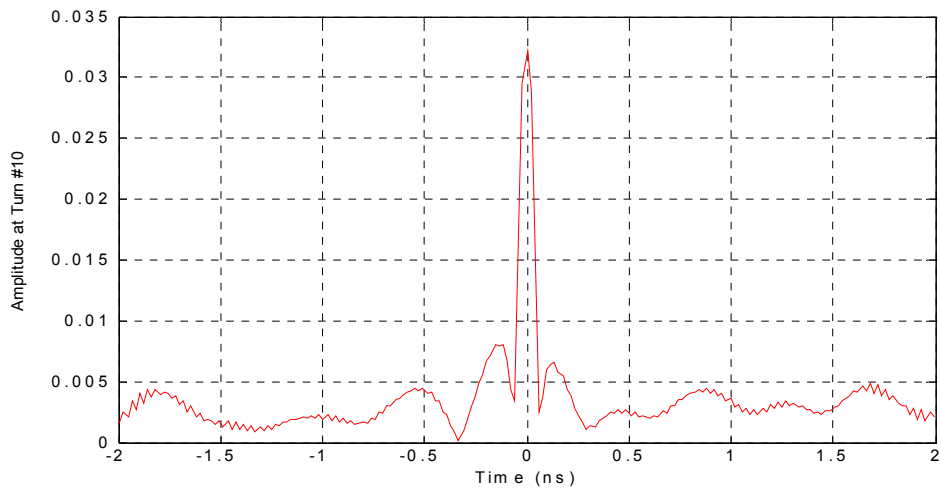

Fig. 1(h)

Fig. 1(a) in the situation of $\sigma$=0.02 ns, the injected MB.

Fig. 1(b) the energy loss per Booster turn due to the real impedance.

Fig. 1(c) phase shifts due to the real impedance, the complex angle, and the total phase shift per Booster turn are shown as the red, black, and blue curves separately.

Fig. 1(d) the RWS of the bunch center at turn #1.

Fig. 1(e) the RWS of the bunch center at turn #2.

Fig. 1(f) the RWS of the bunch center at turn #3.

Fig. 1(g) the RWS of the bunch center at turn #5.

Fig. 1(h) the RWS of the bunch center at turn #10.

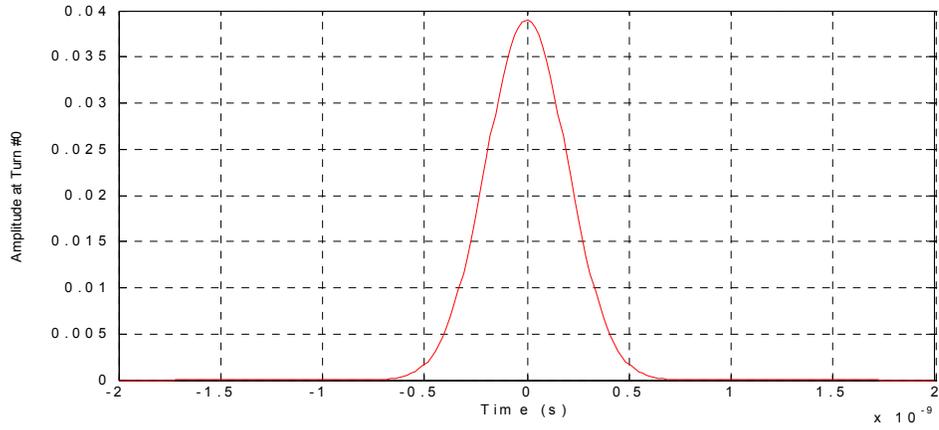

Fig. 2(a)

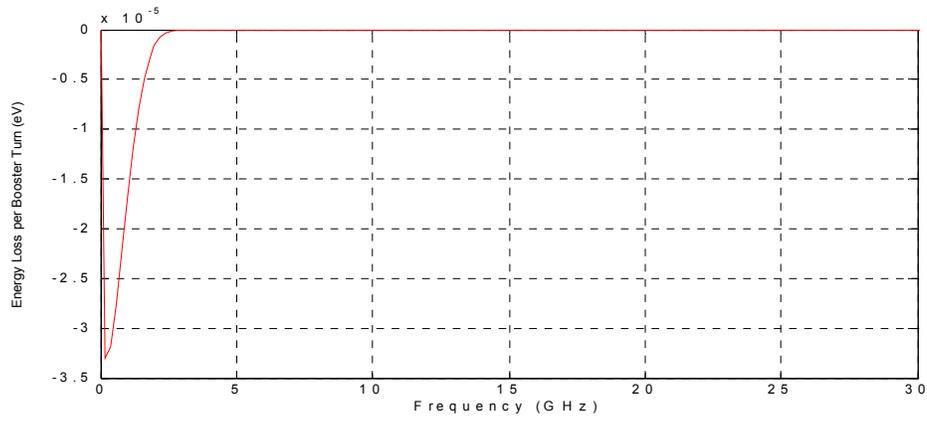

Fig. 2(b)

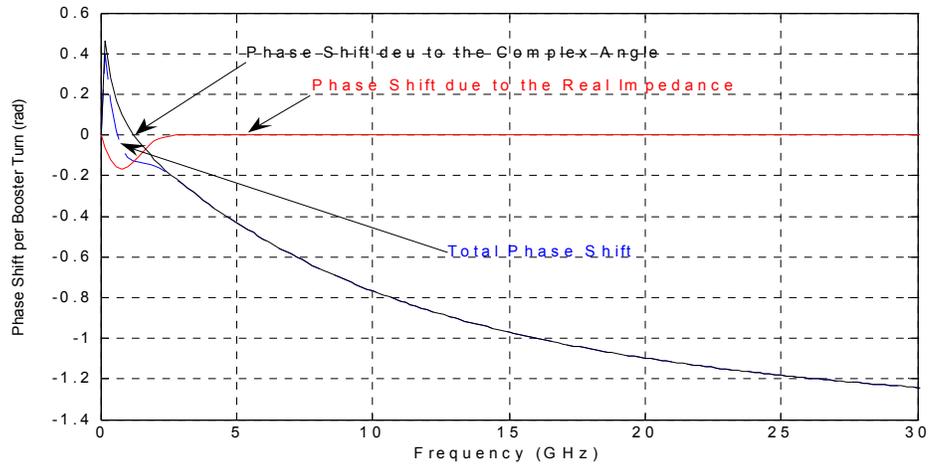

Fig. 2(c)

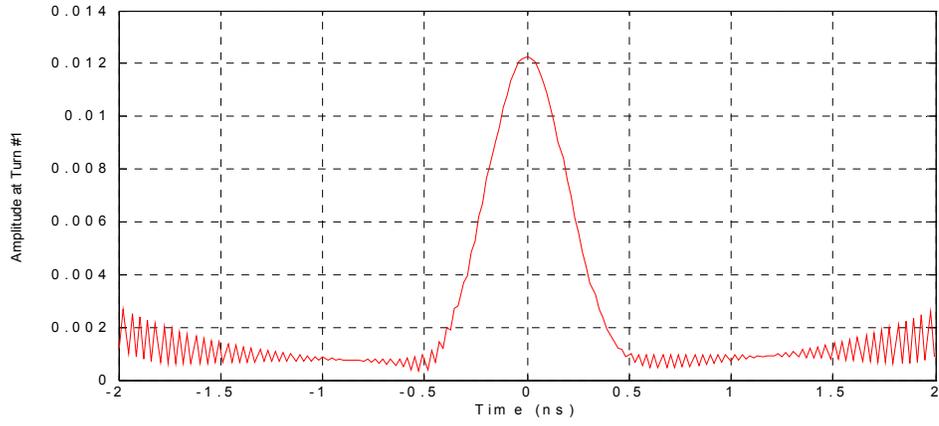

Fig. 2(d)

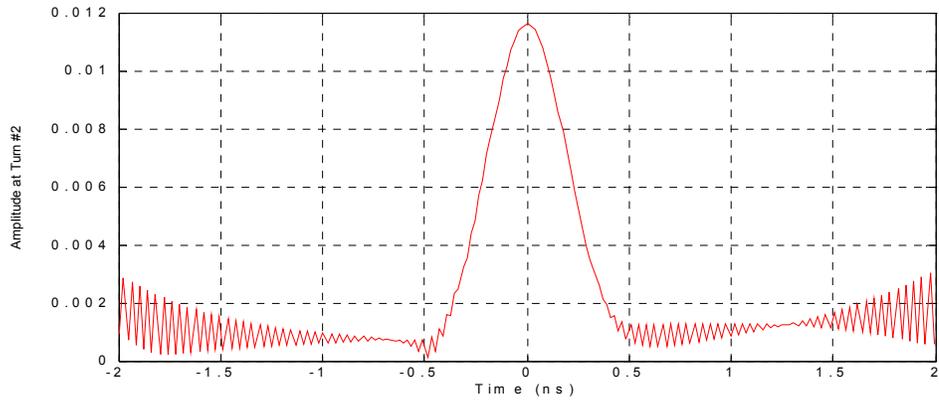

Fig. 2(e)

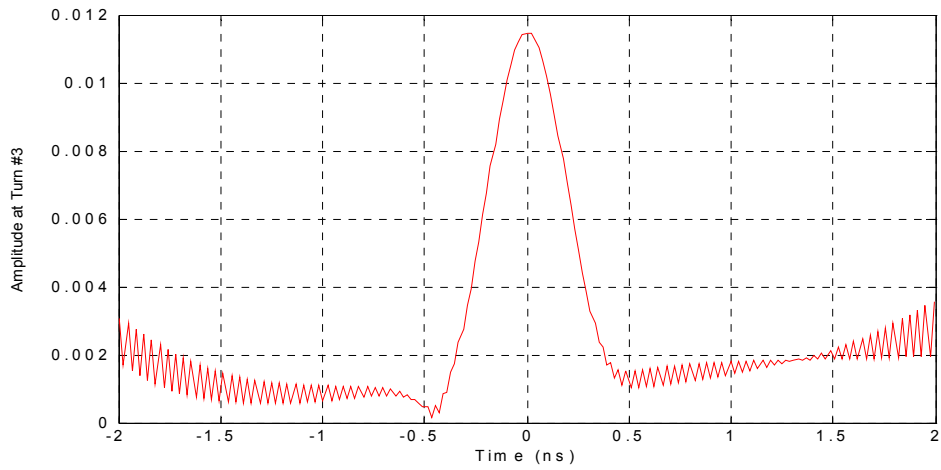

Fig. 2(f)

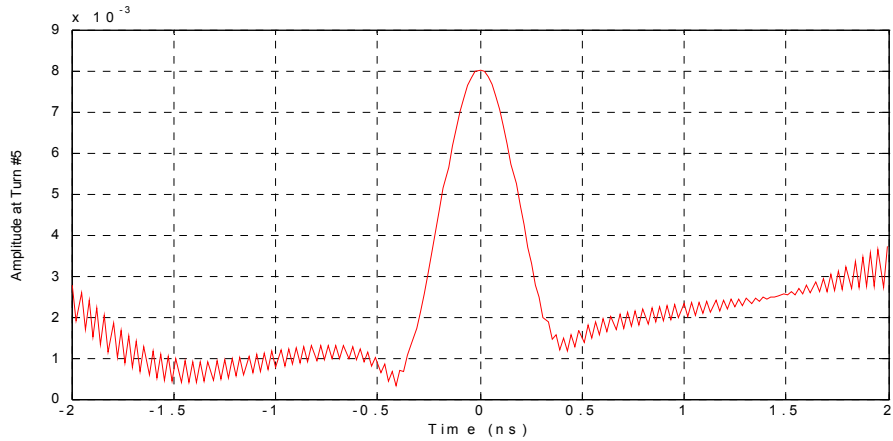

Fig. 2(g)

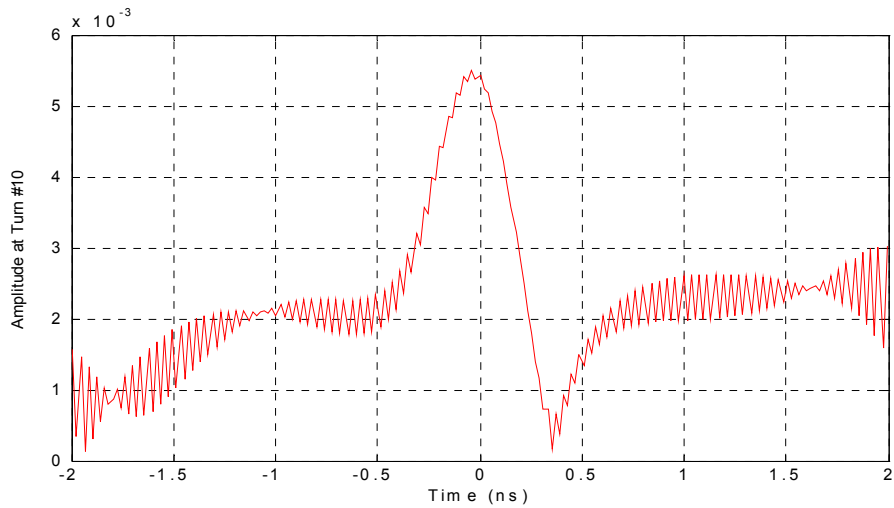

Fig. 2(h)

Fig. 2(a) in the situation of $\sigma$=0.2 ns, the injected MB.

Fig. 2(b) the energy loss per Booster turn due to the real impedance.

Fig. 2(c) phase shifts due to the real impedance, the complex angle, and the total phase shift per Booster turn are shown as the red, black, and blue curves separately.

Fig. 2(d) the RWS of the bunch center at turn #1.

Fig. 2(e) the RWS of the bunch center at turn #2.

Fig. 2(f) the RWS of the bunch center at turn #3.

Fig. 2(g) the RWS of the bunch center at turn #5.

Fig. 2(h) the RWS of the bunch center at turn #10.